\documentclass[12pt]{article}
\newlength{\absize}
\pdfoutput=1
\usepackage{latexsym}
\usepackage{amssymb}
\usepackage{graphicx}
\usepackage{footmisc}
\usepackage{braket}
\renewcommand{\arraystretch}{2}
\newcommand{\tr}{\mathop{\rm Tr}\nolimits}
\usepackage{pictexwd}
\newdimen\tdim
\tdim=\unitlength
\def\stpltsmbl{\setplotsymbol ({\small .})}

\newbox\sru
\setbox\sru=\hbox{\beginpicture
\setcoordinatesystem units <\tdim,\tdim>
\stpltsmbl
\setquadratic
\plot
  0.0   0.0
  4.8   1.5
  7.5   5.0
  7.3   8.5
  5.0  10.0
  2.7   8.5
  2.5   5.0
  5.2   1.5
 10.0   0.0
/
\endpicture}
\def\springru #1 #2 *#3 /{\multiput {\copy\sru}  at
#1 #2 *#3 10 0 /}

\newcommand{\xbar}[1]{\overline{#1}}

\usepackage{hyperref}

\begin{document}

\thispagestyle{empty}
\pagestyle{empty}
\renewcommand{\thefootnote}{\fnsymbol{footnote}}
\newcommand{\starttext}{\newpage\normalsize
 \pagestyle{plain}
 \setlength{\baselineskip}{3ex}\par
 \setcounter{footnote}{0}
 \renewcommand{\thefootnote}{\arabic{footnote}}
 }
\newcommand{\preprint}[1]{\begin{flushright}
 \setlength{\baselineskip}{3ex}#1\end{flushright}}
\renewcommand{\title}[1]{\begin{center}\LARGE
 #1\end{center}\par}
\renewcommand{\author}[1]{\vspace{2ex}{\Large\begin{center}
 \setlength{\baselineskip}{3ex}#1\par\end{center}}}
\renewcommand{\thanks}[1]{\footnote{#1}}
\renewcommand{\abstract}[1]{\vspace{2ex}\normalsize\begin{center}
 \centerline{\bf Abstract}\par\vspace{2ex}\parbox{\absize}{#1
 \setlength{\baselineskip}{2.5ex}\par}
 \end{center}}

\title{Complementarity and Stability Conditions}
\author{
 Howard~Georgi\thanks{\noindent \tt hgeorgi@fas.harvard.edu}
 \\ \medskip
 Center for the Fundamental Laws of Nature\\
 The Physics Laboratories \\
 Harvard University \\
 Cambridge, MA 02138
 }
\centerline{06/16}
\abstract{We discuss the issue of complementarity between the confining
phase and the Higgs phase for gauge theories 
in which there are no light particles
below the scale of confinement or spontaneous symmetry breaking.
We show with a number of examples
that even though the low energy effective theories are the same
(and trivial), discontinuous changes in the stucture of heavy stable
particles can signal a phase transition and 
thus we can sometimes argue that 
two phases which have different structures of heavy
particles that cannot be continuously connected
and thus the phases
cannot be complementary.  We discuss what this means and 
suggest that such
``stability conditions'' can be a useful physical 
check for complementarity.
} 

\starttext

\setcounter{equation}{0}


\section{Introduction}
This note is an attempt to understand better the classic papers 
by Fradkin and Shenker~\cite{Fradkin:1978dv}, Banks and
Rabinovici~\cite{Banks:1979fi}, 't~Hooft~\cite{'tHooft:1979bh} 
and Dimopoulos, Raby and
Susskind~\cite{Susskind:1979up, Raby:1979my} 
related to
complementarity between the Higgs and confining phases in gauge
theories.\footnote{See also \protect{\cite{Wegner:1984qt}}.
One other reference that might be useful is
\protect\cite{Caudy:2007sf}.}   In model building, this is important
because it sometimes
happens that one takes a Higgsed theory that is 
perturbatively
calculable for small couplings and pushes it into regions in which
perturbation theory is questionable.  If the Higgs phase and confining
phase are complementary, that is if there
is no phase transition separating the Higgs phase and confining phase,
then one may hope that this will give a picture of the physics that is
qualitatively correct even if it is not quantitatively reliable.  But if
the two phases are genuinely different, then you have no right to expect
that this procedure will make any sense at all.

A recent example is an $SU(N+3) \times 
SU(3)\times U(1)$ model that was suggested as a possible explanation of the
di-photon exess at $750$~GeV.~\cite{Georgi:2016xhm}  
The model has $(N+3,\xbar 3)$
scalar field $\xi$ that is trying to break the symmetry down to
$SU(N)\times SU(3)\times U(1)$.\footnote{There are no other matter fields
that carry the $SU(N+3)$.}
In the limit in which only one of the couplings gets strong, 
we can think of the strong non-Abelian group as the gauge symmetry
and treat the other approximately as a global symmetry.

If $SU(3)$ gets strong and $SU(N+3)$ is global, the issue is easy.  
Here, I think that there is no hope of complementarity.  Because in this
case, in the Higgs phase, we
have the $SU(N+3)\times U(1)$ global symmetry broken down to 
 $SU(N)\times SU(3)\times U(1)$.  
There is a coset space
\begin{equation}
\frac{SU(N+3)}{SU(N)\times SU(3)}
\end{equation}
describing an $(N,\xbar 3)$ of massless
Goldstone bosons in the Higgs phase and there is no unbroken gauge symmetry
And even if the $SU(N+3)$ is weakly
gauged, the heavy vectors are light and still present in the low energy theory.

In the confining $SU(3)$ theory, there is no reason for the global
$SU(N+3)$ to break and no reason for anything to be light.  So in this
situation, the phases are distinguished by
different symmetries and different massless particles in the low energy theory. 

What happens if $SU(N+3)$
gets strong?  Then presumably the $SU(3)$ is unbroken both in the confining
phase and in the Higgs phase.
So this could perhaps be complementary.
In the Higgs phase we have massless $SU(N)$ gauge bosons, and
the rest of the $SU(N+3)$ gauge bosons have mass of order $gv$.  And
$\Lambda_N$ is of the same order of magnitude times the exponential factor
that goes to 1 as the coupling gets large.  Thus in the gauge invariant
spectrum there are glueballs and bound states of heavy vectors.  
As the coupling increases,  all of these things get heavy!  Likewise, in
the confining phase of the full $SU(N+3)$ theory, we expect that all the
the particle states will have mass of the order of the $SU(N+3)$ confinement
scale or greater.  

Thus in both the confining phase and the Higgs phase, the low energy
theories are trivial. This is consistent with complementarity, and in this
case, we believe that the phases are in fact complementary.  However, in
general, the equivalence of the effective low energy theories
in the confining and Higgs phases~\cite{'tHooft:1979bh}
is not a sufficient condition for complementarity.\footnote{This has been
emphasized in a very different context 
in \protect{\cite{Terning:1997xy}}.}
And we suggest another diagnostic
for complementarity that can be useful.  

It may be that even when the low energy particles and symmetries acting on
them are identical, there are sectors describing 
heavy particles in the two phases with
different properties that distinguish the two phases.  The property that we
will focus on is stability.  In a sense, a heavy stable particle is part of
the effective low energy theory because if something puts one in the
low-energy world, it stays there and its
interactions do not involve any high-energies.~\cite{Georgi:1990um}  
Stability conditions can be
an easy and very physical way of identifying this situation. 

It is important to note that stability for a particular set of parameters
is not enough because complementarity is about how the physics changes as
parameters change. We are interested in the situation in which stability is
guaranteed independent of the phase space.  An example of this is a theory
with a conserved quantized charge.   A conserved charge divides the space
of physical states up into sectors with definite charge, separated by
superselection rules.  
In a theory with a single conserved charge, the sector with the lowest
non-zero positive charge 
must contain stable states - either a single particle with the minimum
charge or a collection of stable particles with total charge equal to the
minimum. 
There is
 stability here, but it is not a property of the particle.  We
can certainly imagine changing the parameters in the theory continuously to
make some a different particle carrying the conserved charge (not necessarily the
same value of the charge) the lightest
particle. And indeed, no single particle with the lowest charge has to
exist at all.  But at least some particles carrying the charge will always be stable so
long as the charge is conserved.  We might say that each sector of charged
states is unconditionally stable, because there is always some combination
of particles that is the lightest state with the appropriate charge.

As a very explicit (and fairly silly) 
example imagine a world with a conserved charge and
three types of  charged particle, $A$, $B$ and $C$ with charges 2, 3, and 5
respectively.  The lowest positive charge is 1, and the stable states in
the charge 1 sector could be $\bar A B$,  $\bar A\bar A C$ or $\bar CBB$, depending
on the particle masses.  Charge conservation guarantees that two of the
particle types are stable, and which two are actually stable depends on the
masses, but the charge 1 sector is is stable independent of the details of
the masses..

If in a phase transition, the lowest positive charge changes, then even if
the light particles in the two phases are qualitatively similar, the
possible structures of stable particles in the effective low energy theory
must be different in the two phases.  There is then no way to get
continuously from one effective theory to the other, and the two phases
cannot be complementary.

In the remainder of this note, we will give a series of examples based on
familiar $SU(N)$ groups. We
hope they will convince the reader that this is an interesting approach.

\section{SU(5) with a scalar 10\label{sec:5-10}}

As a warm-up, and to get the reader used to the style of analysis,
consider an $SU(5)$ theory with a single $10$ of scalars,
$\xi^{jk}=-\xi^{kj}$.  
The most
general renormalizable Lagrangian has a global $U(1)$ symmetry, and for a
range of parameters, $\xi$ develops a VEV that can be put in the
form\footnote{See section~\protect\ref{subsec:5-10}.  Note that this
statement is not trivial, and such details are 
too often 
ignored in treatments of Higgs theories.  However, here, we want to focus
on other things, so in this and subsequent sections, we will relegate the
discussion of the potentials to appendix~\protect\ref{potentials}.
}
{\renewcommand{\arraystretch}{1.2}\begin{equation}
\braket{\xi}=\left(\begin{array}{ccccc}
0&0&0&0&0\\
0&0&0&0&0\\
0&0&0&0&0\\
0&0&0&0&-v\\
0&0&0&v&0
\end{array}\right)
\end{equation}}%
This breaks the $SU(5)$ gauge symmetry down to $SU(3)\times SU(2)$, under which
$\xi$ transforms as
\begin{equation}
(\xbar3,1)+(3,2)+(1,1)
\label{xi32}
\end{equation} 
with the VEV in the $(1,1)$. The $(3,2)$ and the imaginary part of the $(1,1)$ are
eaten by the Higgs mechanism producing a $(3,2)$ and $(1,1)$ of massive
vector bosons.  

There is also a global $U(1)$ symmetry that is
a combination of the original global $U(1)$ and the $U(1)$ generator of the
$SU(5)$ that commutes with $SU(3)\times SU(2)$.  The $(1,1)$ in
(\ref{xi32}) must be neutral under the unbroken symmetry, so the charges
must look like
\begin{equation}
(\xbar3,1)_{2}+(3,2)_1+(1,1)_0
\label{xi32q}
\end{equation}
in some arbitrary normalization.. 
And because the $U(1)$ charge of the multiplet must be the average charge
of the multiplet after symmetry breaking, we know that $\xi$ is a
$10_{6/5}$.  The condensate also breaks the global 5-ality of the $SU(5)$
theory. down to triality$\times$duality for the $SU(3)\times SU(2)$
In the Higgsed theory, the uneaten $(\bar 3,1)$ of scalars has triality 2
and charge $2$, the $(3,2)$ massive gauge boson 
has triality 1, duality 1 and charge $1$.

In both the Higgs phase and the confining phase, heavy particles carry a
quantized conserved charge. 
Now we can examine the stable sectors in the Higgs phase and the
confining phase.  In this case, they match up perfectly.  In the Higgs
phase, all the triality and duality zero 
gauge singlet combinations like 3 $(\xbar3,1)_{2}$ scalars 
or 6 $(3,2)_{1}$ massive vector bosons all
have $U(1)$ charges which are multiples of $6$.  In the confining theory
the 5-ality zero states are
combinations of 5 $10_{6/5}$ scalars, which have the same property.  The
lowest positive charge is $6$ in both cases.   

Thus the
stability conditions do not distinguish between this Higgs phase and the
confining phase, and this is consistent with complementarity.

\section{SU(5) with a scalar 15\label{sec:5-15}}

Contrast the model discussed in section~\ref{sec:5-10}
with an $SU(5)$ theory with a single $15$ of scalars, $\xi^{jk}=\xi^{kj}$.  The most
general renormalizable Lagrangian again has a global $U(1)$ symmetry, and for a
range of parameters, $\xi$ develops a VEV that can be put in the
form\footnote{See section~\protect{\ref{subsec:5-15}}.}
{\renewcommand{\arraystretch}{1.2}\begin{equation}
\braket{\xi}=\left(\begin{array}{ccccc}
0&0&0&0&0\\
0&0&0&0&0\\
0&0&0&0&0\\
0&0&0&0&0\\
0&0&0&0&v
\end{array}\right)
\end{equation}}
This breaks the $SU(5)$ gauge symmetry down to an $SU(4)$, under which
$\xi$ transforms as
\begin{equation}
10+4+1
\label{xi41}
\end{equation} 
with the VEV in the $1$. The $4$ and the imaginary part of the $1$ are
eaten by the Higgs mechanism producing a $4$ and $1$ of massive vector bosons. 

There is also a global $U(1)$ symmetry that is
a combination of the original global $U(1)$ and the $U(1)$ generator of the
$SU(5)$ that commutes with $SU(4)$.  The $1$ in
(\ref{xi41}) must be neutral under the unbroken symmetry, so the charges
must look like (again in an arbitrary normalization)
\begin{equation}
10_{2}+4_1+1_0
\label{xi41q}
\end{equation} 
And because the $U(1)$ charge of the multiplet must be the average charge
of the multiplet after symmetry breaking, we know that $\xi$ is a
$15_{8/5}$.  

This time the heavy stable particle sectors in the Higgs phase and the
confining phase have different $U(1)$ quantum numbers. 
In the Higgs phase, a bound state of 4 $4_1$ massive vector bosons 
confined by the $SU(4)$ has
charge $4$.  The sector with charge 4 has the smallest non-zero value of the
conserved $U(1)$ charge, and thus it is unconditionally stable. 

In the confining phase, there are no states with charge
$4$.  The lowest nonzero charged state is a bound state of 5
$15_{8/5}$ scalars, with charge $8$, and the lightest charge 8 particle is
unconditionally stable.  Thus the Higgs and confining theories have
different unconditionally stable sectors and cannot be complementary.

The Higgs phase and the confining phase are distinguished
in spite of the fact that there is nothing in the low energy theory in
either case, because the stable heavy
particle sectors have different global $U(1)$ charges.  There is no complementarity.

It is interesting to compare this with a model with $\xi$ being a single $5$ of
scalars, where we know that complementarity is preserved.  
In this case, again the gauged $SU(5)$ is
broken to $SU(4)$ preserving a global, but now  $\xi$ breaks up into  
\begin{equation}
4_1+1_0
\label{xi41q-5}
\end{equation}
and again the $4_1$ is eaten by the Higgs mechanism to become the
longitudinal component of the massive gauge boson.  The Higgs phase in this
case is missing the $10_{2}$ of scalars, but otherwise looks remarkably
similar to the $15$ case.  The 4-ality zero states have charges that are
multiples of $4$.  But now the confining phase is not qualitatively
different, because the $5$ has global charge $4/5$ (the average charge of
the mutiplet in (\ref{xi41q-5})), so the 5-ality zero states also have
charges that are multiples of $4$.  

One of the issues in the difference between $\xi=15$ and $\xi=5$ is that the
charge structure of the Higgs phase is determined in part by the charges of
the eaten Goldstone bosons which depend on the symmetry breaking but are
independent of the details of the rest of the $\xi$ multiplet.  But in the
composite phase, the full multiplet is involved in everything.

\section{SU(5) with 3 scalar 10s\label{sec:5-310s}}

  Next consider an
$SU(5)$ gauge group with three $10$s of scalars.
We can write the scalar fields as
\begin{equation}
\xi^{ajk}
=\xi^{a[jk]}
\end{equation}
where $a$ is the
$SU(3)$ flavor index and $j,k$ are $SU(5)$ indices.  We show below that
we can find a 
potential with a global $SU(3)\times U(1)$ symmetry that produces the
vev\footnote{See section~\protect{\ref{subsec:5-310s}}.}
\begin{equation}
\braket{\xi^{ajk}}=v\,\epsilon^{ajk}
\label{1-v}
\end{equation}
where  $\epsilon_{ajk}$ is the
3-dimensional Levi-Civita tensor. 

The VEV (\ref{1-v}) preserves a global $SU(3)$ symmetry generated by the sum of 
the global $SU(3)_G$ symmetry generator and the generator of an
$SU(3)_g$ subgroup
of the gauged $SU(5)$ acting on the first 3 of the $SU(5)$ indices.  And it
preserves a
gauged $SU(2)$ acting on $SU(5)$ indices $4$ and $5$.  Under
$SU(3)_G\times SU(5)_g\to 
SU(3)_{G}\times SU(3)_{g}\times SU(2)_g\to 
SU(3)_{G+g}\times SU(2)_g$, 
the $SU(5)$ generators break up into
\begin{equation}
(1,24)\to (1,1,3)+(1,8,1)+(1,3,2)+(1,\xbar3,2)+(1,1,1)
\to (1,3)+(8,1)+(3,2)+(\xbar3,2)+(1,1)
\end{equation}
and
the (complex) $\xi$s transform like
\begin{equation}
(3,10)
\to
(3,\xbar3,1)+(3,3,2)+(3,1,1)
\to
(8,1)+(1,1)+(6,2)+(\xbar3,2)+(3,1)
\label{1-su4}
\end{equation}
The vev  (\ref{1-v}) is in the real part of the singlet.
The imaginary part of the $(8,1)$ and $(1,1)$ in (\ref{1-su4}) and the
$(\xbar3,2)$ in (\ref{1-su4}) 
are eaten by the Higgs mechanism
giving massive gauge
bosons, producing an $SU(3)$ adjoint, a complex $(3,2)$ and a singlet.
If the gauge coupling is small, their 
masses are in the ratio
$1:1:\sqrt{8/5}$.   But the details here don't really matter if the coupling is strong.
They just all get heavy.

There is also a global $U(1)$ symmetry that is
a combination of the original global $U(1)$ and the $U(1)$ generator of the
$SU(5)$ that commutes with $SU(3)\times SU(2)$.  
The $SU(3)_{G+g}$ singlet in $\xi$ must be neutral under the unbroken
$U(1)$, so the charges must look like
\begin{equation}
(3,\xbar3,1)_0+(3,3,2)_q+(3,1,1)_{2q}
\to
(8,1)_0+(1,1)_0+(6,2)_q+(\xbar3,2)_q+(3,1)_{2q}
\label{1-su4q}
\end{equation}
for some $q$.
The global charge of the $\xi$ field in the theory before symmetry breaking
must then be $4q/5$, because this is the average charge of the multiplet 
after symmetry.

In the Higgs phase, there are no states with charge $q$, so 
the sector with charge $2q$ must contain stable particles with total charge
$2q$.  
For weak coupling, 
there are both ``fundamental'' charge $2q$ states, like the scalar
$(3,1)_{2q}$ in (\ref{1-su4q}), and composite charge $2q$ states, like the
bound states of two $(\xbar3,2)_q$ vector bosons, confined when the unbroken
$SU(2)$ gauge interaction gets strong.

In the confining phase, on
the other hand, physical states confined by the strong $SU(5)$ gauge
interactions must have 5-ality 0.  They therefore
contain a multiple of 
$5$ $\xi$s, and thus have charges which are a multiple of $4q$ and there is
no stable sector with charge $2q$.  Thus the
Higgs phase defined 
by (\ref{1-v}) cannot not complementary to the confining phase.

Note that the global $SU(3)$ symmetry here is almost certainly not
necessary.  It makes the analysis of the potential much easier, but if it
is explicity broken, a phase with the same unbroken $U(1)$ and the same
charges will very likely
exist in some region of the parameter space.

\section{SU(5) with 4 scalar {\boldmath$\xbar{10}\,$}s\label{sec:5-410bars}}

The examples in sections~\ref{sec:5-15} and
\ref{sec:5-310s} have confining unbroken 
gauge symmetries in the Higgs phase.  Again, this is not necessary.
Here 
is an example similar to that in section~\ref{sec:5-310s}, but slightly
more complicated in which the non-Abelian 
gauge symmetry is completely broken.
Consider an
$SU(5)$ gauge group with four $\xbar{10}\,$s of scalars.
We can write the scalar fields as
\begin{equation}
\xi^{ajk\ell}
=\xi^{a[jk\ell]}
\end{equation}
where $a$ is the
$SU(4)$ flavor index and $j,k,\ell$ are $SU(5)$ indices.  Here we can find a 
potential with a global $SU(4)\times U(1)$ symmetry that produces the
vev\footnote{See section~\protect{\ref{subsec:5-410bars}}.}
\begin{equation}
\braket{\xi^{ajk\ell}}=v\,\epsilon^{ajk\ell}
\label{2-v}
\end{equation}
where  $\epsilon_{ajk\ell}$ is the
4-dimensional Levi-Civita tensor. 
As usual, we first discuss the non-Abelian structure and then go
back and discuss the $U(1)$s.

The VEV (\ref{2-v}) preserves a global $SU(4)$ symmetry generated by the sum of 
the global $SU(4)_G$ symmetry generator and the generator of an
$SU(4)_g$ subgroup
of the gauged $SU(5)$ acting on the first 4 of the $SU(5)$ indices.  Under
$SU(4)_G\times SU(4)_g\to SU(4)_{G+g}$, the (complex) $\xi$s transform like
\begin{equation}
(4,6)+(4,\xbar4)\to 20+\xbar4+15+1
\label{2-su4}
\end{equation}
When the singlet gets a vev corresponding to  (\ref{1-v}), the $SU(5)$
symmetry breaks completely and
the $SU(5)$ generators break up into
\begin{equation}
24\to 15+4+\xbar4+1
\end{equation}
all of which eat parts of the $\xi$ field giving rise to massive gauge bosons.
At tree level, this gives mass to all the gauge
bosons, producing an $SU(4)$ adjoint, a $4+\xbar4$ and $1$ with masses in the ratio
$1:\sqrt{3/2}:3/\sqrt{5}$.   Again the details here don't really matter
if the coupling is strong.

As in the example in section~\ref{sec:5-310s},
there is also a global $U(1)$ symmetry that is
a combination of the original global $U(1)$ and the $U(1)$ generator of the
$SU(5)$.  
The $SU(4)_{G+g}$ singlet in $\xi$ must be neutral under the unbroken
$U(1)$, so the charges must look like
\begin{equation}
(4,6)_q+(4,\xbar4)_0\to 20_q+\xbar4_q+15_0+1_0
\label{2-su4q}
\end{equation}
for some $q$.
The global charge of the $\xi$ field in the theory before symmetry breaking
is the average charge of the multiplet which is $3q/5$.

Now the Higgs phase at small coupling 
has particles with charge are the $q$ --- for
example
the $\xbar4$
state.
In the confining phase, however, the physical states are all built out of
multiples of $5$
$\xi$s and thus have charges which are multiples of $3q$.

So again, in this case, this Higgs phase and the confining phase are distinguished
in spite of the fact that there is nothing in the low energy theory in
either case, because there are different stable sectors of heavy particles. As in
section~\ref{sec:5-310s}, the $SU(4)$ global symmetry makes it easy to
analyze the more general potential, but it is probably not necessary for
the stability analysis, which depends only on the global $U(1)$.

\section{Conclusion}
The examples in this note should convince the reader that in constructing
an effective theory, it is important to consider heavy stable particles as
well as light particles.  This can contain important information about the
structure of the quantum field theory.  In particular, we have shown that 
discontinuous changes in the stucture of heavy stable
particles can signal a phase transition.  While this can show conclusively
that two phases are not continuously related, we do not know of any way to
sharpen these argument to determine conclusively that two phases are
complementary.  For this we still need ``theorems'' like those of
reference~\cite{Fradkin:1978dv} and \cite{Banks:1979fi}.

\section{Acknowledgements}
Savas Dimopoulos, 
Yuichiro Nakai, Stuart Raby, Matt Reece, Matt Schwartz, Steve Shenker and 
Lenny Susskind have
contributed with important 
remarks. HG is supported in part by the National
Science Foundation under grant PHY-1418114.

\appendix
\section{Potentials and 
VEVs\label{potentials}}

\setcounter{subsection}{1}
\subsection{SU(5) with a scalar 10\label{subsec:5-10}}
For an $SU(5)$ theory with a single $10$ of scalars,
$\xi^{jk}=-\xi^{kj}$,  
we want to show that the most
general renormalizable Lagrangian has a global $U(1)$ symmetry, and for a
range of parameters, $\xi$ develops a VEV that can be put in the form
{\renewcommand{\arraystretch}{1.2}\begin{equation}
\braket{\xi}=\left(\begin{array}{ccccc}
0&0&0&0&0\\
0&0&0&0&0\\
0&0&0&0&0\\
0&0&0&0&-v\\
0&0&0&v&0
\end{array}\right)
\label{a1vev}
\end{equation}}%
which breaks the symmetry down to $SU(3)\times SU(2)$.
This is easy because we can treat the $\xi$ field as a $2\times2$ matrix and
write the most general renormalizable potential as
\begin{equation}
\lambda_1\left(\Big(\tr(\xi\xi^\dagger)\Bigr)^2-4v^2\tr(\xi\xi^\dagger)\right)
-\lambda_2\left(\tr(\xi\xi^\dagger\xi\xi^\dagger)
-2v^2\tr(\xi\xi^\dagger)\right)
\end{equation}
This evidently has a global $U(1)$ and it
is extremized for the VEV (\ref{a1vev}).  
If 
\begin{equation}
2\lambda_1>\lambda_2>0
\end{equation}
then (\ref{a1vev}) is a local minimum.  The massive scalars are a $(1,1)$
with mass squared 
$8 \left(2 \lambda _1-\lambda _2\right) v^2$ and a complex $(\xbar3,1)$
with mass squared $4\lambda_2v^2$.

\subsection{SU(5) with a scalar 15\label{subsec:5-15}}
For an $SU(5)$ theory with a single $15$ of scalars,
$\xi^{jk}=\xi^{kj}$,  
we want to show that the most
general renormalizable Lagrangian has a global $U(1)$ symmetry, and for a
range of parameters, $\xi$ develops a VEV that can be put in the form
{\renewcommand{\arraystretch}{1.2}\begin{equation}
\braket{\xi}=\left(\begin{array}{ccccc}
0&0&0&0&0\\
0&0&0&0&0\\
0&0&0&0&0\\
0&0&0&0&0\\
0&0&0&0&v
\end{array}\right)
\label{a2vev}
\end{equation}}%
which breaks the symmetry down to $SU(4)$.
Again we can treat the $\xi$ field as a $2\times2$ matrix and
this time we will
write the most general renormalizable potential as
\begin{equation}
\lambda_1\left(\Big(\tr(\xi\xi^\dagger)\Bigr)^2-2v^2\tr(\xi\xi^\dagger)\right)
-\lambda_2\left(\tr(\xi\xi^\dagger\xi\xi^\dagger)
-2v^2\tr(\xi\xi^\dagger)\right)
\end{equation}
This again has a global $U(1)$ and it
is extremized for the VEV (\ref{a2vev}).  
If 
\begin{equation}
\lambda_1>\lambda_2>0
\end{equation}
then (\ref{a2vev}) is a local minimum.  The massive scalars are a real singlet
with mass squared 
$8 \left( \lambda _1-\lambda _2\right) v^2$ and a complex $10$
with mass squared $4\lambda_2v^2$.

\subsection{SU(5) with 3 scalar 10s\label{subsec:5-310s}}

Here we are interested an
$SU(5)$ gauge group with three $10$s of scalars
which we write as
\begin{equation}
\xi^{ajk}
=-\xi^{akj}
\end{equation}
where $a$ is the
$SU(3)$ flavor index and $j,k$ are $SU(5)$ indices.  We show below that
we can find a 
potential with a global $SU(3)\times U(1)$ symmetry that produces the vev
\begin{equation}
\braket{\xi^{ajk}}=v\,\epsilon^{ajk}
\label{a41-v}
\end{equation}
where  $\epsilon_{ajk}$ is the
3-dimensional Levi-Civita tensor.

The VEV (\ref{a41-v}) preserves a global $SU(3)$ symmetry generated by the sum of 
the global $SU(3)_G$ symmetry generator and the generator of an
$SU(3)_g$ subgroup
of the gauged $SU(5)$ acting on the first 3 of the $SU(5)$ indices.  And it
preserves a
gauged $SU(2)$ acting on $SU(5)$ indices $4$ and $5$.

To see that this Higgs phase actually exists, consider the most general potential.
The potential must involve 2 $\xi$s and 2 $\xi^\dagger$s.  Bose symmetry
implies that the 2 $\xi$ transform like
\begin{equation}
(3,10)\times(3,10)_{\rm symmetric}
=(6,5)+(6,50)+(\xbar{3},45)
\label{a41-bose}
\end{equation}
so there are three independent quartic terms in the potential which we can
take to be
\begin{equation}
\kappa_1=\kappa_0^2
\quad\mbox{where $\kappa_0$ is the invariant mass term}\quad
\kappa_0=
\xi^{aj_1k_1}
\,
\xbar\xi_{aj_1k_1}
\end{equation}
\begin{equation}
\kappa_2=\xi^{bj_1k_1}
\,
\xbar\xi_{aj_1k_1}
\,
\xi^{aj_2k_2}
\,
\xbar\xi_{bj_2k_2}
\end{equation}
\begin{equation}
\kappa_3=\xi^{aj_2k_1}
\,
\xbar\xi_{aj_1k_1}
\,
\xi^{bj_1k_2}
\,
\xbar\xi_{bj_2k_2}
\end{equation}
If we then write the most general
potential as
\begin{equation}
V=\lambda_1(\kappa_1-12v^2\kappa_0)
+\lambda_2(\kappa_2-4v^2\kappa_0)
-\lambda_3(\kappa_3-4v^2\kappa_0)
\end{equation}
$V$ is extremized for the vev (\ref{1-v}), and if the $\lambda$s satisfy
\begin{equation}
3\lambda_1+\lambda_2>\lambda_3\;,\quad
4\lambda_2>\lambda_3\;,\quad
\lambda_3>0
\end{equation} 
then (\ref{a41-v}) is a local minimum so the example works.
The squared masses of the massive scalars are
{\renewcommand{\arraystretch}{1.3}
\begin{equation}
\begin{array}{cc}
 \mbox{a real $(1,1)$} & 16 v^2 \left(3 \lambda _1+\lambda _2-\lambda
   _3\right) \\
 \mbox{a real $(8,1)$} & 4 v^2 \left(4 \lambda _2-\lambda _3\right) \\
 \mbox{a complex $(6,2)$} & 4 v^2 \lambda _3 \\
 \mbox{a complex $(3,1)$} & 8 v^2 \lambda _3 \\
\end{array}
\end{equation}}

\subsection{SU(5) with 4 scalar {\boldmath$\xbar{10}\,$}s\label{subsec:5-410bars}}
The examples in sections~\ref{sec:5-15} and
\ref{sec:5-310s} have confining unbroken 
gauge symmetries in the Higgs phase.  Again, this is not necessary.
Here 
is an example similar to that in section~\ref{sec:5-310s}, but slightly
more complicated example in which the non-Abelian 
gauge symmetry is completely broken.
Again consider an
$SU(5)$ gauge group with four $\xbar{10}\,$s of scalars.
We can write the scalar fields as
\begin{equation}
\xi^{ajk\ell}
=\xi^{a[jk\ell]}
\end{equation}
where $a$ is the
$SU(4)$ flavor index and $j,k,\ell$ are $SU(5)$ indices.  Here we can find a 
potential with a global $SU(4)\times U(1)$ symmetry that produces the vev
\begin{equation}
\braket{\xi^{ajk\ell}}=v\,\epsilon^{ajk\ell}
\label{a52-v}
\end{equation}
where  $\epsilon_{ajk\ell}$ is the
4-dimensional Levi-Civita tensor.

The VEV (\ref{a52-v}) preserves a global $SU(4)$ symmetry generated by the sum of 
the global $SU(4)_G$ symmetry generator and the generator of an
$SU(4)_g$ subgroup
of the gauged $SU(5)$ acting on the first 4 of the $SU(5)$ indices.  Under
$SU(4)_G\times SU(4)_g\to SU(4)_{G+g}$, the (complex) $\xi$s transform like

We can analyze the potential as we did in the previous example.
The potential must involve two $\xi$s and 2 $\xi^\dagger$s.  Bose symmetry
implies that the two $\xi$ transform like
\begin{equation}
(4,\xbar{10})\times(4,\xbar{10})_{\rm symmetric}
=(10,5)+(10,\xbar{50})+(6,\xbar{45})
\label{a5bose}
\end{equation}
so there are three independent quartic terms in the potential which we can
take to be
\begin{equation}
\kappa_1=\kappa_0^2
\quad\mbox{where $\kappa_0$ is the invariant mass term}\quad
\kappa_0=
\xi^{aj_1k_1\ell_1}
\,
\xbar\xi_{aj_1k_1\ell_1}
\end{equation}
\begin{equation}
\kappa_2=\xi^{bj_1k_1\ell_1}
\,
\xbar\xi_{aj_1k_1\ell_1}
\,
\xi^{aj_2k_2\ell_2}
\,
\xbar\xi_{bj_2k_2\ell_2}
\end{equation}
\begin{equation}
\kappa_3=\xi^{aj_2k_1\ell_1}
\,
\xbar\xi_{aj_1k_1\ell_1}
\,
\xi^{bj_1k_2\ell_2}
\,
\xbar\xi_{bj_2k_2\ell_2}
\end{equation}
We could write down a 4th along the same lines, 
\begin{equation}
\xi^{bj_2k_1\ell_1}
\,
\xbar\xi_{aj_1k_1\ell_1}
\,
\xi^{aj_1k_2\ell_2}
\,
\xbar\xi_{bj_2k_2\ell_2}
\end{equation}
but we know from
(\ref{a5bose}) that it is not independent.  
If we then write the most general
potential as
\begin{equation}
V=\lambda_1(\kappa_1-48v^2\kappa_0)
+\lambda_2(\kappa_2-12v^2\kappa_0)
-\lambda_3(\kappa_3-12v^2\kappa_0)
\end{equation}
Then $V$ is extremized for the vev (\ref{a52-v}), and if the $\lambda$s satisfy
\begin{equation}
4\lambda_1+\lambda_2>\lambda_3\;,\quad
9\lambda_2>\lambda_3\;,\quad
\lambda_3>0
\end{equation} 
then (\ref{a52-v}) is a local minimum.
The squared masses of the massive scalars are
\begin{equation}
\begin{array}{cc}
 \mbox{a real singlet} & 48 v^2 \left(4 \lambda _1+\lambda
 _2-\lambda_3\right) \\ 
 \mbox{a real $15$} & \frac{16}{3} v^2 \left(9 \lambda _2-\lambda _3\right) \\
 \mbox{a complex $20$} & 8 v^2 \lambda _3 
\end{array}
\end{equation}

\bibliography{large-g}

\end{document}